\documentclass{PoS}
\usepackage{physics}
\usepackage{url}
\usepackage{comment}
\usepackage{simplewick}

\def\simge{\mathrel{%
       \rlap{\raise 0.511ex \hbox{$>$}}{\lower 0.511ex \hbox{$\sim$}}}}
\def\simle{\mathrel{
       \rlap{\raise 0.511ex \hbox{$<$}}{\lower 0.511ex \hbox{$\sim$}}}}

\newcommand{\Slash}[1]{{\ooalign{\hfil/\hfil\crcr\(#1\)}}}

\title{Measuring of chiral susceptibility using gradient flow}

\ShortTitle{Measuring of chiral susceptibility using gradient flow}

\author{\speaker{Atsushi Baba} , Asobu Suzuki\\
        Graduate School of Pure and Applied Sciences, University of Tsukuba, Tsukuba, Ibaraki,  305-8571, Japan\\
        E-mail: \email{ababa@het.ph.tsukuba.ac.jp}}
        
\author{Shinji Ejiri\\
        Department of Physics, Niigata University, Niigata 950-2181, Japan}
        
\author{Kazuyuki Kanaya\\
        Tomonaga Center for the History of the Universe, University of Tsukuba, Tsukuba, Ibaraki,  305-8571, Japan}
        
\author{Masakiyo Kitazawa\\
        Department of Physics, Osaka University, Osaka 560-0043, Japan\\
        J-PARC Branch, KEK Theory Center, Institute of Particle and Nuclear Studies, KEK, 203-1, Shirakata, Tokai, Ibaraki, 319-1106, Japan} 

\author{Takanori Shimojo, Hiroshi Suzuki\\
        Department of Physics, Kyushu University, 744 Motooka, Nishi-ku, Fukuoka 819-0395, Japan}

\author{Yusuke Taniguchi\\
        Center for Computational Science (CCS), University of Tsukuba, Tsukuba, Ibaraki,  305-8571, Japan}

\author{Takashi Umeda\\
        Graduate School of Education, Hiroshima University, Higashihiroshima, Hiroshima 739-8524, Japan}

\abstract{In lattice QCD with Wilson-type quarks, the chiral symmetry is explicitly broken by the Wilson term on finite lattices. Though the symmetry is guaranteed to recover in the continuum limit, a series of non-trivial procedures are required to recover the correct renormalized theory in the continuum limit.
Recently, a new use of the gradient flow technique was proposed, in which correctly renormalized quantities are evaluated in the vanishing flow-time limit.
This enables us to directly study the chiral condensate and its susceptibility with Wilson-type quarks.
Extending our previous study of the chiral condensate and its disconnected susceptibility in (2+1)-flavor QCD 
at a heavy $u$, $d$ quark mass ($m_{\pi}/m_{\rho}\simeq0.63$) and approximately physical $s$ quark mass,
we compute the connected contributions to the chiral susceptibility in the temperature range of 178--348 MeV
on a fine lattice with $a\simeq0.07$ fm. \\
Preprint numbers: UTHEP-731, UTCCS-P-121, KYUSHU-HET-192, J-PARC-TH-0154
}

\FullConference{The 36th Annual International Symposium on Lattice Field Theory - LATTICE2018\\
		22-28 July, 2018\\
		Michigan State University, East Lansing, Michigan, USA.}

\begin{document}
\section{Introduction}
In the QCD transition/crossover between the low-temperature hadron phase and high-temperature Quark--Gluon Plasma (QGP) phase, the chiral symmetry plays an essential role.
The spontaneously broken chiral symmetry at low temperatures recovers at this transition. 
The order parameter, the chiral condensate $\ev{\bar{\psi}_f \psi_f}$ for the $f$'th flavor, and its susceptibility 
\begin{align}
\chi^{\mathrm{full}}_{f} = \ev{ \left\{ \frac{1}{N_{\Gamma}} \sum_x \bar{\psi}_f(x) \psi_f(x) \right\}^2} 
- \ev{ \frac{1}{N_{\Gamma}} \sum_x \bar{\psi}_f(x) \psi_f(x) }^2 ,
\end{align}
where $N_{\Gamma} = \sum_x 1$ is the lattice volume and $f$ is not summed over in the RHS,
are thus among the most basic observables to detect the transition/crossover. 
The chiral susceptibility can be further decomposed into connected and disconnected parts as 
$ \chi^{\mathrm{full}}_{f} =  \chi^{\mathrm{conn}}_f + \chi^{\mathrm{disc}}_{f}$, where
\begin{eqnarray}
\chi^{\mathrm{conn}}_f &=& 
\ev{  \frac{1}{N_{\Gamma}} \sum_x 
\contraction[1.5ex]{}{\bar{\psi}}{{}_f(x) \psi_f(x) \bar{\psi}_f(0)}{\psi}
\contraction{\bar{\psi}_f(x)}{\psi}{{}_f(x)}{\bar{\psi}} 
\bar{\psi}_f(x) \psi_f(x) \bar{\psi}_f(0) \psi_f(0) } , 
\label{eq:conn}
\\
\chi^{\mathrm{disc}}_f &=& 
\ev{ \left\{ \frac{1}{N_{\Gamma}} \sum_x 
\contraction{}{\bar{\psi}}{{}_f(x)}{\psi}
\bar{\psi}_f(x) \psi_f(x) \right\}^2 } 
- \ev{ \frac{1}{N_{\Gamma}} \sum_x \bar{\psi}_f(x) \psi_f(x) }^2 .
\label{eq:disc}
\end{eqnarray}

Because the QCD transition/crossover is a non-perturbative phenomenon, a lattice study is called for.
On the lattice, however, the chiral symmetry is not an easy property to be realized. 
With Wilson-type quarks, for example, the chiral symmetry is explicitly broken by the Wilson term to remove the doublers.
Although the chiral symmetry is guaranteed to be recovered in the continuum limit, non-trivial renormalization procedures including additive renormalization to fermion masses, chiral condensation etc.\ are required.
With staggered-type quarks, though a subgroup of the chiral symmetry is kept on the lattice, lattice artifacts due to the taste degrees of freedom thus introduced should be removed carefully.
With lattice chiral quarks such as the domain wall quark and the overlap quark, a lattice-modified version of the chiral symmetry, the Ginsparg-Wilson symmetry, can be realized at finite lattice spacings in a limit of additional parameters.
In compensation for this, however, typically hundreds times more computational resources are required.

Recently, a new use of the gradient flow method \cite{Narayanan:2006rf,GF1,GF2,GF3} was proposed to calculate correctly renormalized observables from lattice simulations \cite{suzuki1,suzuki2}:
Making use of the finiteness of flowed operators, non-perturbative estimates of observables are extracted by taking a vanishing flow-time extrapolation.
The new method was first applied to the energy-momentum tensor 
for which the explicit violation of the Poincar\'e invariance on the lattice has been a hard obstacle in obtaining a non-perturbative estimate \cite{suzuki1,EMT1,EMT2}.
However, because the method is quite general, we can apply it also to obtain correctly renormalized chiral observables from lattice simulations \cite{suzuki2}.

In Refs.~\cite{EMT2,TS2}, we have applied the method to study quantities including the chiral condensate, disconnected chiral susceptibility, and topological susceptibility in $(2+1)$-flavor full QCD at finite temperature, adopting improved Wilson quark action and Iwasaki gauge action.
The results we obtained are quite encouraging. 
We see clear signals of the QCD transition/crossover with the chiral condensate and the disconnected chiral susceptibility, even with the explicit chiral violation of Wilson-type quarks \cite{EMT2}.
We also find a good consistency between the gluonic and fermionic definitions for the topological susceptibility \cite{TS2}.
In this paper, we extend the study of Ref.~\cite{EMT2} to compute the connected contribution (\ref{eq:conn}) of the chiral susceptibility.

\section{Lattice setup}

We study $(2 + 1)$-flavor QCD adopting a non-perturbatively $O(a)$-improved Wilson quark action and the RG-improved Iwasaki gauge action.
We apply the fixed-scale approach, i.e.\ we vary the temperature $T = 1 / a N_t$ by varying the temporal lattice size $N_t$ at a fixed simulation point. 
This enables us to use a common set of zero-temperature configurations to carry out the zero-temperature subtraction of thermal observables and also the fermion wave function renormalization (see Sec.~\ref{sec:gf}) at all temperatures,
thus reducing the cost of zero-temperature simulations \cite{fixed_scale}.
As the zero-temperature configuration, 
we have chosen a set of CP-PACS+JLQCD configurations generated at $\beta = 2.05$ corresponding to $a \simeq 0.07$ fm, degenerate $u$, $d$ quark mass corresponding to $m_{\pi}/m_{\rho} \simeq 0.63$, and almost physical $s$ quark mass corresponding to $m_{\eta_{ss}}/m_{\phi} \simeq 0.74$ \cite{zero_temp_conf}.
The bare PCAC quark masses are $am_{ud} = 0.02105(17)$ and $am_{s} = 0.03524(26)$.

At this simulation point, finite temperature configurations have been generated on lattices with $N_t=4$, 6, $\cdots$ $16$ corresponding to the tempereture range of 174--697 MeV \cite{nonzero_temp_conf}. 
In our previous study applying the gradient flow method to these configurations \cite{EMT2}, however, it turned out that the lattices with $N_t \simle 8$ suffer from sizable $O\left((aT)^2=1/N_t^2\right)$ errors. 
Therefore, in this paper, we concentrate on the range $N_t \ge 8$ corresponding to $T\simeq178$--348 MeV, as listed in Table~\ref{tab:parameters}.

\begin{table}
\centering
\caption{Lattice parameters}
\label{tab:parameters}
\begin{tabular}{c|c|c|c|c}
\hline
 $N_s$ & $N_t$ & $T$ [MeV] & $T / T_c$ & configuration \\ \hline \hline
28 & 56 & 0 & 0 & 65 \\
32 & 16 & 174 & 0.92 & 144 \\
32 & 14 & 199 & 1.05 & 127 \\
32 & 12 & 232 & 1.22 & 129 \\
32 & 10 & 279 & 1.47 & 78 \\
32 & 8 & 348 & 2.44 & 51 \\
\hline
\end{tabular}
\end{table}

\section{Gradient flow method}
\label{sec:gf}

We adopt the simplest gradient flow for the gauge field \cite{GF1}:
\begin{align} \label{eq:GF_gauge}
\partial_{t} B_{\mu} (t, x) = D_{\nu} G_{\nu \mu} (t, x) ,\qquad B_{\mu} (0, x) = A_{\mu} (x),
\end{align}
where the field strength $G_{\nu \mu}$ and the covariant derivative $D_{\nu}$ are defined in terms of the flowed gauge field $B_{\mu}$.
%
The flow for quarks are given by~\cite{GF3}
\begin{gather}\label{eq:GF_fermion}
\partial_{t} \chi_f (t, x) = D^2 \chi_f (t, x) ,\qquad \chi_f (0, x) = \psi_f(x) , \\
\partial_{t} \bar{\chi}_f (t, x) = \bar{\chi}_f (t, x) \overleftarrow{D} \,^2 ,\qquad \bar{\chi}_f (0, x) = \bar{\psi}_f (x) ,
\end{gather}
with $D_{\mu} \chi_f (t, x) = \left( \partial_{\mu} + B_{\mu} (t, x) \right) \chi_f$ and
$\bar{\chi}_f (t, x) \overleftarrow{D}_\mu = \bar{\chi}_f (t, x) \left( \overleftarrow{\partial_{\mu}} - B_{\mu} (t, x) \right)$.

In terms of the flowed fields,
the correctly normalized chiral condensate in the $\overline{\textrm{MS}}$ scheme at $\mu=2$ GeV is given by~\cite{suzuki2}
\begin{align} \label{eq:condensate}
\ev{\bar{\psi}(x)_f \psi_f(x)} &= \lim _{t \rightarrow 0} \, c_s (t)  \varphi_f (t) \ev{ \bar{\chi}_f(t, x) \chi_f(t, x)},
\end{align}
where the matting coefficient $c_s (t)$ and fermion wave function renormalization factor $\varphi_f (t)$ are
\begin{gather}
c_s (t) = \left\{ 1 + \frac{\bar{g}^2(1/\sqrt{8t})}{(4 \pi)^2} \left[ 4 \left( \gamma - 2 \ln 2 \right) + 8 + \frac{4}{3} \ln 432 \right] \right\} \frac{\bar{m}_f (1 / \sqrt{8t})}{\bar{m}_f(2 \mathrm{GeV})}, 
\\
\varphi_f (t) = \frac{-6}{ ( 4\pi)^2 t^2 \expval{ \bar{\chi}(t, x) \overset{\leftrightarrow}{\Slash{D}} \chi(t, x) }_0 } .
\end{gather}
Here, $\bar{g}(\mu)$ and $\bar{m}(\mu)$ are running coupling and running mass in the $\overline{\textrm{MS}}$ scheme at scale $\mu$, respectively, and $\ev{\cdots}_0$ and $\ev{\cdots}$ mean zero- and finite-temperature expectation values.
The chiral susceptibility $\chi^{\mathrm{full}}_f$ is given similarly by
\begin{eqnarray}
\chi^{\mathrm{full}}_f 
&=& \lim_{t \rightarrow 0}\, \left[c_s (t)  \varphi_f (t)\right]^2 \left\{ \ev{ \left[ \frac{1}{N_{\Gamma}} \sum_x  \bar{\chi}_f(t, x) \chi_f(t, x) \right]^2} - \ev{ \frac{1}{N_{\Gamma}} \sum_x \bar{\chi}_f(t, x) \chi_f(t, x)}^2 \right\}
\nonumber \\  \label{eq:susceptibility}
&=& \lim_{t \rightarrow 0}\,  \chi^{\mathrm{full}}_f(t) .
\end{eqnarray}

We evaluate (\ref{eq:condensate}) and (\ref{eq:susceptibility}) non-perturbatively by performing lattice simulations.
The proper way to apply the gradient flow method is to take the continuum limit $a \to 0$ first, then the leading small-$t$ correction to the flowed chiral susceptibility will be
$ 
\chi^{\mathrm{full}}_f (t) = \chi^{\mathrm{full}}_f + t A + O(t^{2}), 
$ 
where 
$A$ is the contamination of dimension eight operators.
In our study, however, we have only one lattice spacing so far and want to take the continuum limit later.
To the leading order of $O(a^2)$, 
we will have additional contaminations like 
\begin{align}\label{eq:lattice_artifact}
\chi^{\mathrm{full}}_f (t, a) = \chi^{\mathrm{full}}_f (t) + O( a^{2} / t,\; a^{2} T^{2}\!,\; a^{2} m^{2}\!,\; a^{2} \Lambda_{\mathrm{QCD}}^{2}).
\end{align}
Among the $O(a^2)$ terms, the term $O(a^{2} / t)$ is singular in the $t \to 0$ extrapolation.
In Refs.~\cite{EMT2,TS2}, we have avoided the difficulty by identifying a range of $t$,``linear window'', a range of t in which terms like $O(a^{2} / t)$ and $O(t^2)$ are not dominating.
Taking a linear $t\to0$ extrapolation using the data in the linear window, we may evaluate the RHS's of (\ref{eq:condensate}) and (\ref{eq:susceptibility}) 
up to $O(a^{2} T^{2}\!,\; a^{2} m^{2}\!,\; a^{2} \Lambda_{\mathrm{QCD}}^{2})$ lattice artifacts.
We may check the validity of the linear windows by performing non-linear fits including $O(a^{2} / t)$ and $O(t^2)$ terms.
The difference between the linear and non-linear fits gives an estimate of the systematic error due to the fit ansatz.
See Ref.~\cite{EMT2} for more details.

\section{Numerical results}

Extending the study of Ref.~\cite{EMT2} in which the chiral condensate $\ev{\bar{\psi}_f \psi_f}$ and the disconnected chiral susceptibility (\ref{eq:disc}) are evaluated by the gradient flow method,
we compute the connected contribution of the chiral susceptibility, by calculating quark correlation functions at finite flow time as required in Eq.~(\ref{eq:conn}).

Previous studies of the connected contribution to the chiral susceptibility adopting the conventional method suggest that the connected susceptibility shows only a mild or no response around the QCD transition/crossover:
With $(2+1)$-flavors of HISQ staggered quarks at $m_\pi\sim80$--160 MeV, the connected part varies monotonically around the transition temperature with ``no contribution to the singular behavior'' in the full susceptibility \cite{full_susceptibility_ks}.
With two flavors of overlap chiral quarks at $m_\pi \sim 500$ MeV, they ``do not see a pronounced peak'' in the transition region suggested by the Polyakov loop and the chiral condensate \cite{full_susceptibility_overlap}.
A study on an anisotropic lattice with $(2+1)$-flavors of improved Wilson quarks at $m_\pi = 384(4)$ MeV also found that the bare connected susceptibility shows ``no singular behavior'' around the transition region \cite{full_susceptibility_wilson}.

\begin{figure}[tb]
\begin{minipage}{0.33\hsize}
\centering
\includegraphics[width=\linewidth]{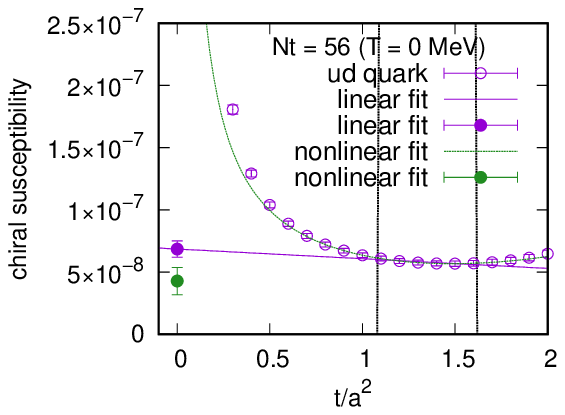}
\end{minipage}
\begin{minipage}{0.33\hsize}
\centering
\includegraphics[width=\linewidth]{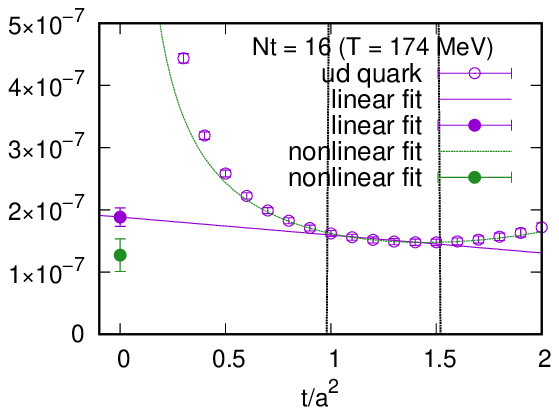}
\end{minipage}
\begin{minipage}{0.33\hsize}
\centering
\includegraphics[width=\linewidth]{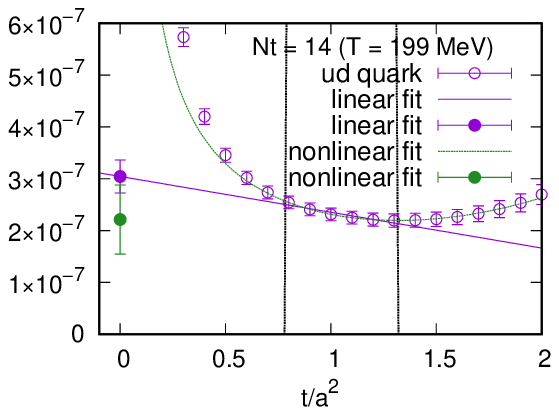}
\end{minipage}
\begin{minipage}{0.33\hsize}
\centering
\includegraphics[width=\linewidth]{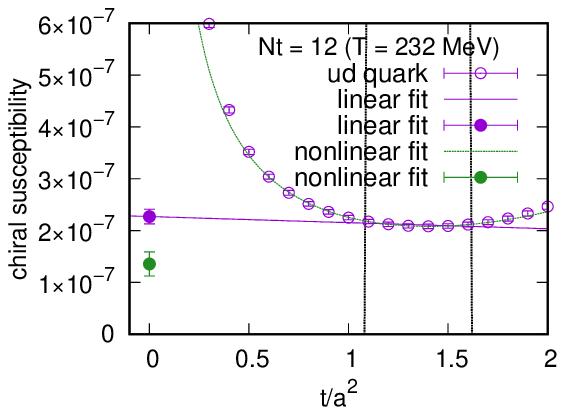}
\end{minipage}
\begin{minipage}{0.33\hsize}
\centering
\includegraphics[width=\linewidth]{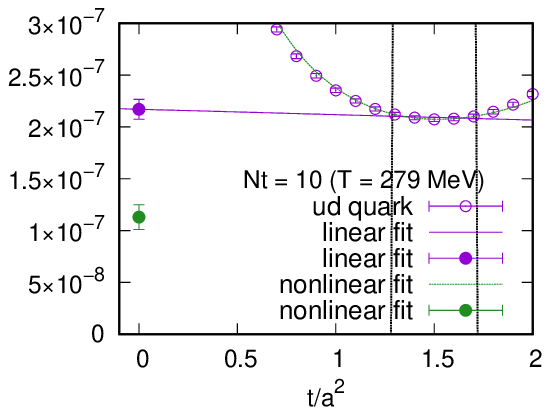}
\end{minipage}
\begin{minipage}{0.33\hsize}
\centering
\includegraphics[width=\linewidth]{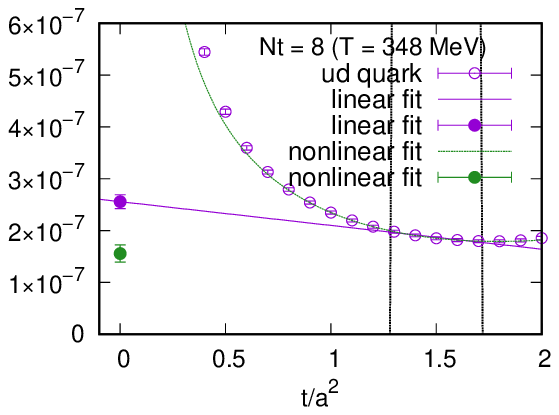}
\end{minipage}
\vspace{-5mm}
\caption{Full chiral susceptibility $\chi^{\mathrm{full}}_{\mathit{ud}}(t)$ as a function of flow time. $\chi^{\mathrm{full}}_{\mathit{ud}}(t)$ is fitted by linear and non linear function. We take $t \to 0$ extrapolation with linear fit and estimate the systematic error by nonlinear fit. The vertical axes are in lattice unit.}
\label{fig:ud_nonlinear_fit}
\end{figure}

With the gradient flow method, we compute correctly renormalized chiral susceptibility (up to $O(a^{2} T^{2}\!,\; a^{2} m^{2}\!,\; a^{2} \Lambda_{\mathrm{QCD}}^{2})$ lattice artifacts) including the connected contributions.
In Fig.~\ref{fig:ud_nonlinear_fit}, we plot our full chiral susceptibility $\chi^{\mathrm{full}}_{\mathit{ud}}(t)$ for the light $u$ or $d$ quark evaluated at finite flow times.
Results for the $s$ quark are similar.
Unlike the case of the energy-momentum tensor \cite{EMT2}, we could not see a clear linear window for the $t \to 0$ extrapolation.
In this paper, we thus try both linear and non-linear fits with varying the fit window.
Here, the non-linear fit means
$ 
\chi^{\mathrm{full}}_f (t, a) = \chi^{\mathrm{full}}_f + t A + t^2 B + \frac{a^2}{t} C
$ 
as adopted in Ref.~\cite{EMT2}.
In this paper, as a trial with the full susceptibility, we adopt linear windows in which $O(a^{2} / t)$ and $O(t^2)$ contributions are minimized in the sense that 
the difference in $\chi^{\mathrm{full}}_f$ is the smallest between the linear and nonlinear fits for the same range of linear window when taking the $t \to 0$ limit.
The resulting fits with the linear windows are shown in Fig.~\ref{fig:ud_nonlinear_fit}.
We adopt the results of the linear fits as central values and take the difference between the two fits as an estimate of the systematic error due to the fit ansatz.

\begin{figure}[tb]
\centering
\begin{minipage}{0.4\hsize}
\centering
\includegraphics[width=\linewidth]{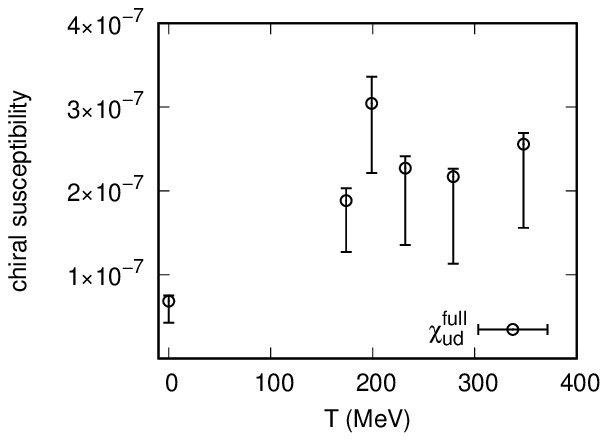}
\end{minipage}
\hspace{0.03\hsize}
\begin{minipage}{0.4\hsize}
\centering
\includegraphics[width=\linewidth]{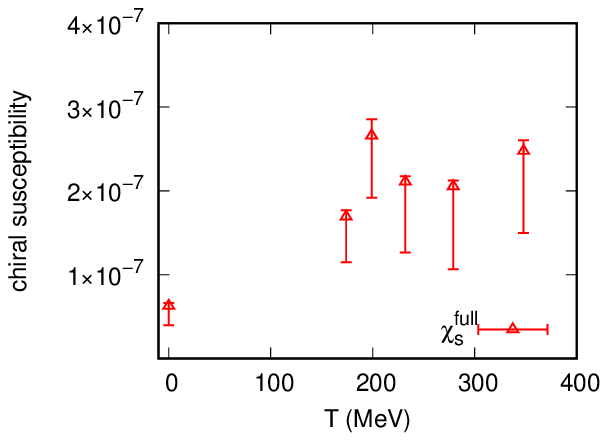}
\end{minipage}
\vspace{-1mm}
\caption{Full chiral susceptibility $\chi^{\mathrm{full}}_f$ as function of the temperature.
The vertical axes are in lattice unit.
Left: light $u$ or $d$ quark susceptibility. Right: $s$ quark susceptibility. }
\label{fig:chi_full}
\end{figure}

The results of the full chiral susceptibility in the $t\to0$ limit are summarized in Fig.~\ref{fig:chi_full} for the $u$ or $d$ quark (left panel) and for the $s$ quark (right).
We see a slight peak at $T=199$ MeV, in accordance with an estimation of $T_{\textrm{pc}}\sim190$ MeV from other observables \cite{nonzero_temp_conf}.

\begin{figure}[tb]
\centering
\begin{minipage}{0.38\hsize}
\centering
\includegraphics[width=\linewidth]{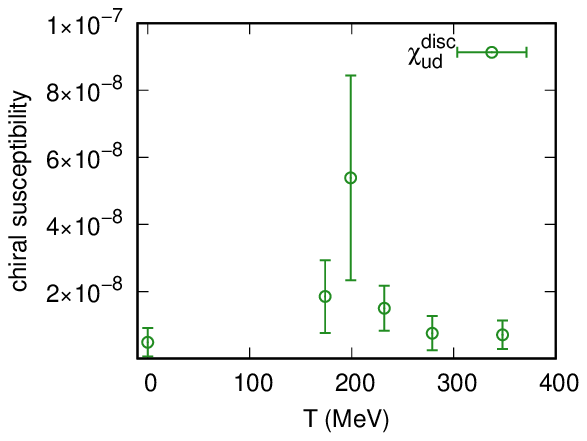}
\end{minipage}
\hspace{0.03\hsize}
\begin{minipage}{0.38\hsize}
\centering
\includegraphics[width=\linewidth]{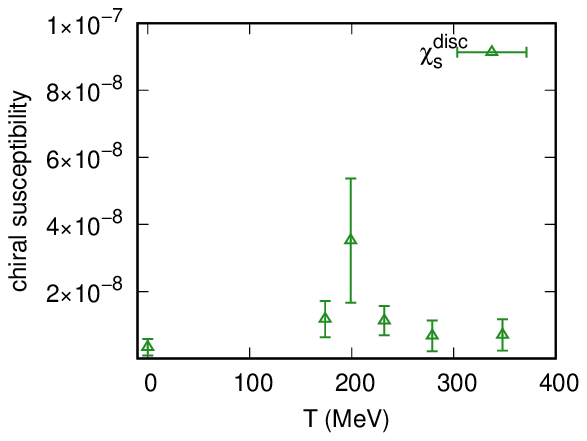}
\end{minipage}
\vspace{-1mm}
\caption{Disconnected chiral susceptibility $\chi^{\mathrm{disc}}_f$ as a function of temperature, adopting the same linear windows as adopted for the full chiral susceptibility.
Left: light $u$ or $d$ quark susceptibility. Right: $s$ quark susceptibility.}
\label{fig:chi_disc}
\end{figure}

\begin{figure}[tb]
\centering
\begin{minipage}{0.38\hsize}
\centering
\includegraphics[width=\linewidth]{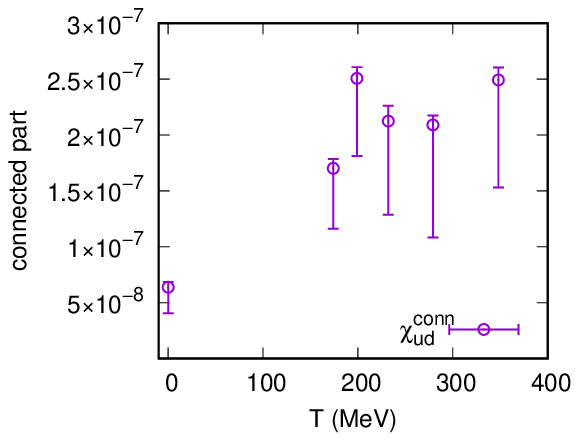}
\end{minipage}
\hspace{0.03\hsize}
\begin{minipage}{0.38\hsize}
\centering
\includegraphics[width=\linewidth]{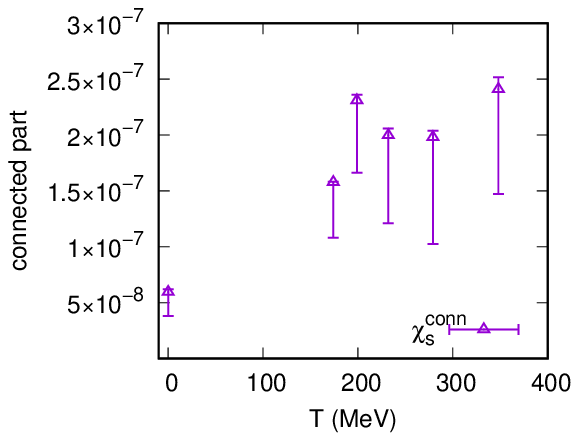}
\end{minipage}
\vspace{-1mm}
\caption{The same as Fig.~\ref{fig:chi_disc} but for the connected chiral susceptibility $\chi^{\mathrm{conn}}_f$.}
\label{fig:connected}
\end{figure}

The results for the disconnected and connected chiral susceptibilities adopting the same linear windows are shown in Figs.~\ref{fig:chi_disc} and \ref{fig:connected}, respectively.
Note that the vertical scales of the plots are different between the disconnected and connected chiral susceptibilities.
As the amount, the connected susceptibility gives about ten times larger contributions to the full susceptibility than the disconnected one. 

As noted in Ref.~\cite{EMT2}, the disconnected chiral susceptibility shows a clear pear at $T=199$ MeV.
On the other hand, the connected chiral susceptibility shows at most a slight bump at $T=199$ MeV, though the errors are still large to draw a definite conclusion. 
We note that the temperature dependence of the connected chiral susceptibility is similar to that observed with overlap chiral quarks \cite{full_susceptibility_overlap}.

\section{Conclusion}
We studied the chiral susceptibility in lattice QCD with $(2 + 1)$-flavors of dynamical Wilson quarks.
We resolve the issue of explicit chiral violation due to the Wilson term by adopting the gradient flow method.
At a heavy $u$, $d$ quark mass ($m_{\pi}/m_{\rho}\simeq0.63$) and approximately physical $s$ quark mass,
we calculated both connected and disconnected chiral susceptibilities in the temperature range of 178--348 MeV
on a fine lattice with $a\simeq0.07$ fm.

We found that the full chiral susceptibility shows a slight peak at $T=199$ MeV, in accordance with previous estimation of $T_{\textrm{pc}}\sim190$ MeV from other observables.
The peak structure is clear with the disconnected chiral condensate, as noted in Ref.~\cite{EMT2}.
Though the errors are large, our connected chiral susceptibility also suggests a mild peak. 
The mild sensitivity of the connected susceptibility to the QCD transition/crossover is in accordance with previous observations using conventional methods.
However, more statistics and more works on the fits are needed to draw a definite conclusion.
In parallel with this, we are reducing the light quark mass down to the physical point, where a sharper chiral transition was suggested by the light quark chiral condensate \cite{physicalpoint}.

\vspace{2mm}
This work was in part supported by JSPS KAKENHI Grant Numbers JP18K03607, JP17K05442, JP16H03982,  JP15K05041, JP26400251, JP26400244, and JP26287040.
This research used computational resources of COMA and Oakforest-PACS provided by the Interdisciplinary Computational Science Program of Center for Computational Sciences, University of Tsukuba,
Oakforest-PACS at JCAHPC through the HPCI System Research Project (Project ID:hp17208), OCTOPUS at Cybermedia Center, Osaka University, and ITO at R.I.I.T., Kyushu University.
The simulations were in part based on Lattice QCD common code Bridge++ \cite{bridge}.


\end{document}